\documentclass{ws-ijmpe}
\usepackage[super,compress]{cite}
\usepackage{hyperref}
\usepackage{subcaption}
\begin{document}
\markboth{Prottoy Das on behalf of ALICE Collaboration}{Hot QCD Matter}
\catchline{}{}{}{}{}
\title{Measurement of charged-particle jet properties in p--Pb collisions at $\sqrt{s_{\rm NN}} = 5.02$ TeV with ALICE}
\author{Prottoy Das on behalf of ALICE Collaboration}
\address{Department of Physics, Bose Institute, Kolkata 700091, West Bengal, India\\ prottoy.das@cern.ch}
\maketitle
%
\begin{abstract}
	We present the measurement of charged-particle jet properties in minimum bias p--Pb collisions at $\sqrt{s_{\rm NN}}$ = 5.02 TeV in ALICE. Jets are reconstructed from charged particles at midrapidity using the anti-$k_{\rm T}$ jet finding algorithm with jet resolution parameter $R$ = 0.4. The mean charged particle multiplicity and jet fragmentation function for leading charged-particle jets in the $p_{\rm T}$ interval 10 $< p_{\rm T, jet}^{\rm ch} <$ 100 GeV/\textit{c} are measured  and compared with theoretical model predictions.	
\end{abstract}


\section{Introduction}
	Jets are collimated sprays of particles produced from the fragmentation and hadronization of hard-scattered partons in high-energy hadronic and nuclear collisions. Jet properties are sensitive to the details of parton showering process and are expected to be modified in the presence of a dense partonic medium. Measurements of intra-jet properties in p--Pb collisions are useful to investigate cold nuclear matter effects~\cite{CNM} and enrich our current understanding of particle production in such collision systems. In this work, we present the measurement of charged-particle jet properties, such as the mean charged particle multiplicity and fragmentation functions, for leading jets in the $p_{\rm T}$ interval 10 $< p_{\rm T, jet}^{\rm ch} <$ 100 GeV/\textit{c} at midrapidity in p--Pb collisions at a center of mass energy per nucleon-nucleon pair $\sqrt{s_{\rm NN}}$ = 5.02 TeV with ALICE. Results are compared with theoretical model predictions.
	
\section{Analysis details}
The data presented here were collected with the ALICE apparatus in 2016. Detail information about the ALICE detector can be found in Ref.~\cite{ALICE_det}. Events are selected for this analysis using a minimum bias trigger condition which requires the coincidence of signals in the V0A and V0C forward scintillator arrays~\cite{V0}. Only events with a primary vertex within 10 cm from the nominal interaction point along the beam direction ($|z_{\rm vertex}|$=0) are considered resulting into 5.15$\times$10$^8$ events. Charged particles reconstructed with the Inner Tracking System (ITS)~\cite{ITS} and the Time Projection Chamber (TPC)~\cite{TPC} are used for the reconstruction of the primary vertex and jets. These detectors are placed inside a large solenoidal magnet that provides a homogeneous magnetic field $B$ = 0.5 T. 

Charged tracks with $p_{\rm T} >$ 0.15 GeV/$c$ within a pseudorapidity range $|\eta| <$ 0.9 over the full azimuth are accepted. Charged-particle jets are reconstructed from the selected tracks using anti-$k_{\rm T}$ jet finding algorithm~\cite{antikT} with the $p_{\rm T}$-recombination scheme of FastJet package~\cite{FastJet} with jet resolution parameter, $R$ = 0.4. Only leading jets (jet of highest $p_{\rm T}$ in an event) with 10 $< p_{\rm T, jet}^{\rm ch} <$ 100 GeV/\textit{c} are considered for this analysis. The mean charged particle multiplicity in leading jet ($\langle N_{\rm ch} \rangle$) and leading-jet fragmentation function ($z^{\rm ch} = p_{\rm T, track}/p_{\rm T,jet}^{\rm ch}$, where $p_{\rm T, track}$ is the $p_{\rm T}$~of jet constituent inside the leading-jet cone) are studied as a function of jet $p_{\rm T}$.

\begin{figure}
	\begin{center}		
		\includegraphics[width=0.9\linewidth]{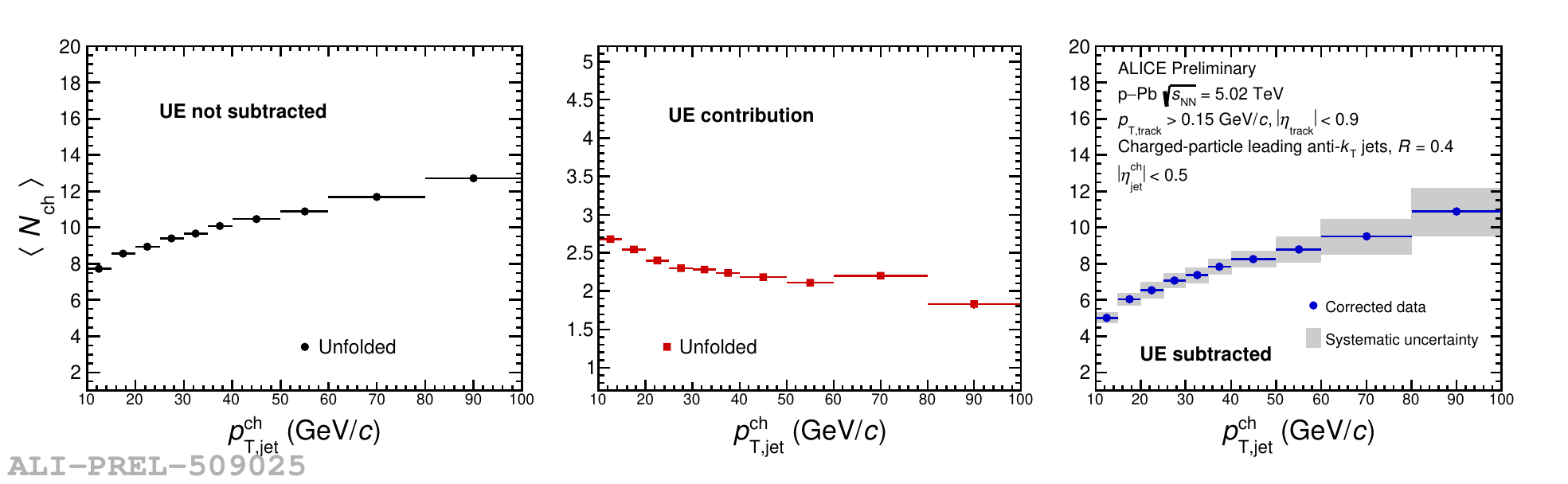}
		\includegraphics[width=0.9\linewidth]{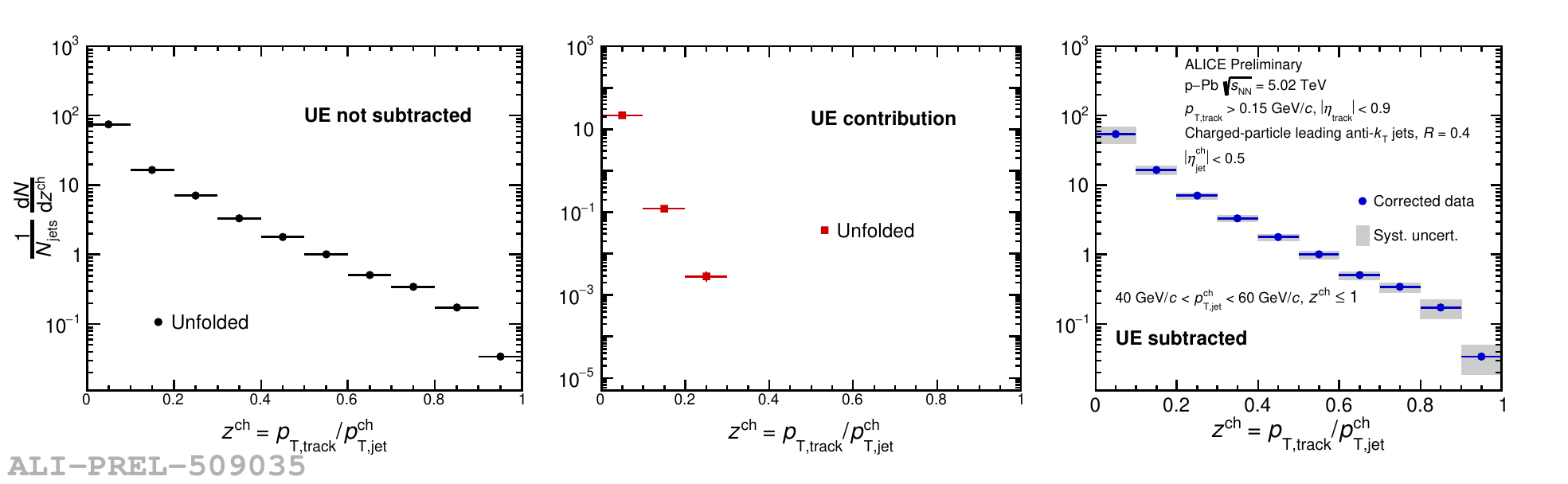}
		\caption{Correction procedure to subtract UE contributions for $\langle N_{\rm ch}\rangle$ and $z^{\rm ch}$ (40 $< p_{\rm T, jet}^{\rm ch} <$ 60 GeV/\textit{c}): Unfolded distributions without UE subtraction (left), UE contribution (middle) and after UE subtraction (right).
		}\label{Fig:UECorrMethod}
	\end{center}
\end{figure}

Contribution from the underlying event (UE; coming from sources other than the hard scattered partons) is estimated using the perpendicular-cone method and subtracted on a statistical basis after unfolding as reported in Fig.~\ref{Fig:UECorrMethod}. In the perpendicular-cone method, circular cones of radius $R$ = 0.4 at the same $\eta$ as the leading jet and perpendicular to the leading jet axis are used for the estimation of the contribution from the UE.

A 2-dimensional Bayesian unfolding technique~\cite{Bayesian} (implemented in RooUnfold~\cite{RooUnfold} package) is applied to correct for instrumental effects such as track-reconstruction efficiency and momentum resolution. For each of the jet observables, a 4D response matrix is constructed from a Monte Carlo (MC) simulation performed with the DPMJET~\cite{DPMJET} event generator and the generated particles are transported through the GEANT detector simulation. DPMJET is a multipurpose generator based on the Dual Parton Model (DPM) and is able to simulate a wide variety of collision systems for energies ranging from a few GeV up to the highest cosmic-ray energies. To construct the response matrix, the detector- and particle-level jets are matched geometrically and only the leading detector-level jet and the corresponding matched particle-level jet in an event are considered. 

Systematic uncertainties from various sources such as tracking efficiency, modelling of the jet properties and the detector response in the MC simulation, choice of regularization parameter or number of iterations in Bayesian unfolding, change in prior distribution, and underlying event contribution are estimated and added in quadrature to calculate the total systematic uncertainty. The total systematic uncertainty is found to be 5\%~-~12\% in the $\langle N_{\rm ch} \rangle$ analysis whereas in the $z^{\rm ch}$ analysis, it is estimated to be $\sim$20\% for 20 $< p_{\rm T, jet}^{\rm ch} <$ 30 GeV/\textit{c} and 25\%~-~45\% for 40 $< p_{\rm T, jet}^{\rm ch} <$ 60 GeV/\textit{c}.

\section{Results and discussion}
\begin{figure}
	\begin{subfigure}[b]{0.4\linewidth}
		\includegraphics[width=0.9\columnwidth]{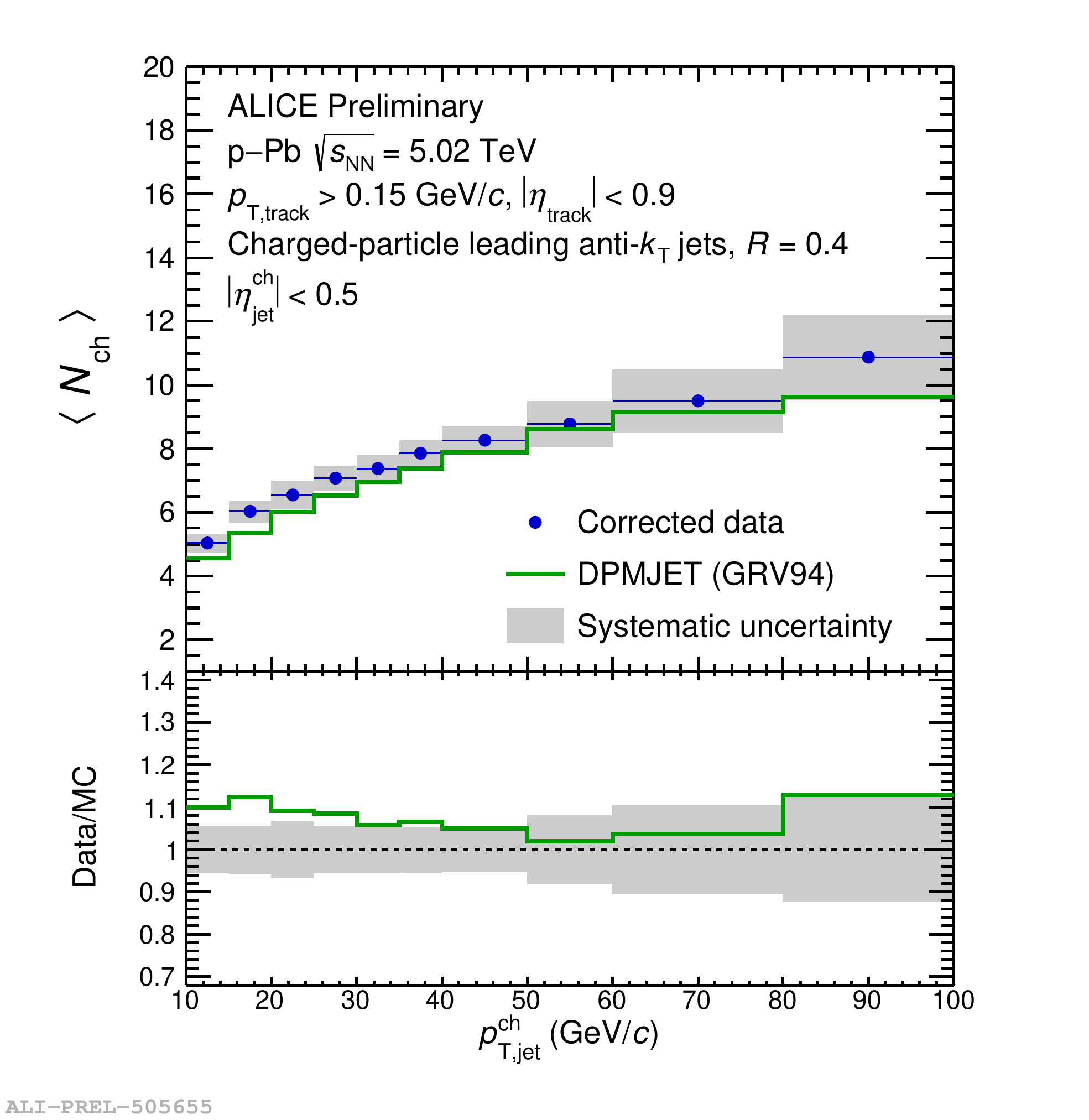}
		\caption{Top panel: Unfolded and UE subtracted $\langle N_{\rm ch}\rangle$ distribution as a function of leading jet $p_{\rm T, jet}^{\rm ch}$. Bottom panel: ratio between the data and MC results.}
		\label{Fig:NchMB}
	\end{subfigure}
	\vspace*{-0.27cm}
	\begin{subfigure}[b]{0.4\linewidth}
		\includegraphics[width=0.9\columnwidth]{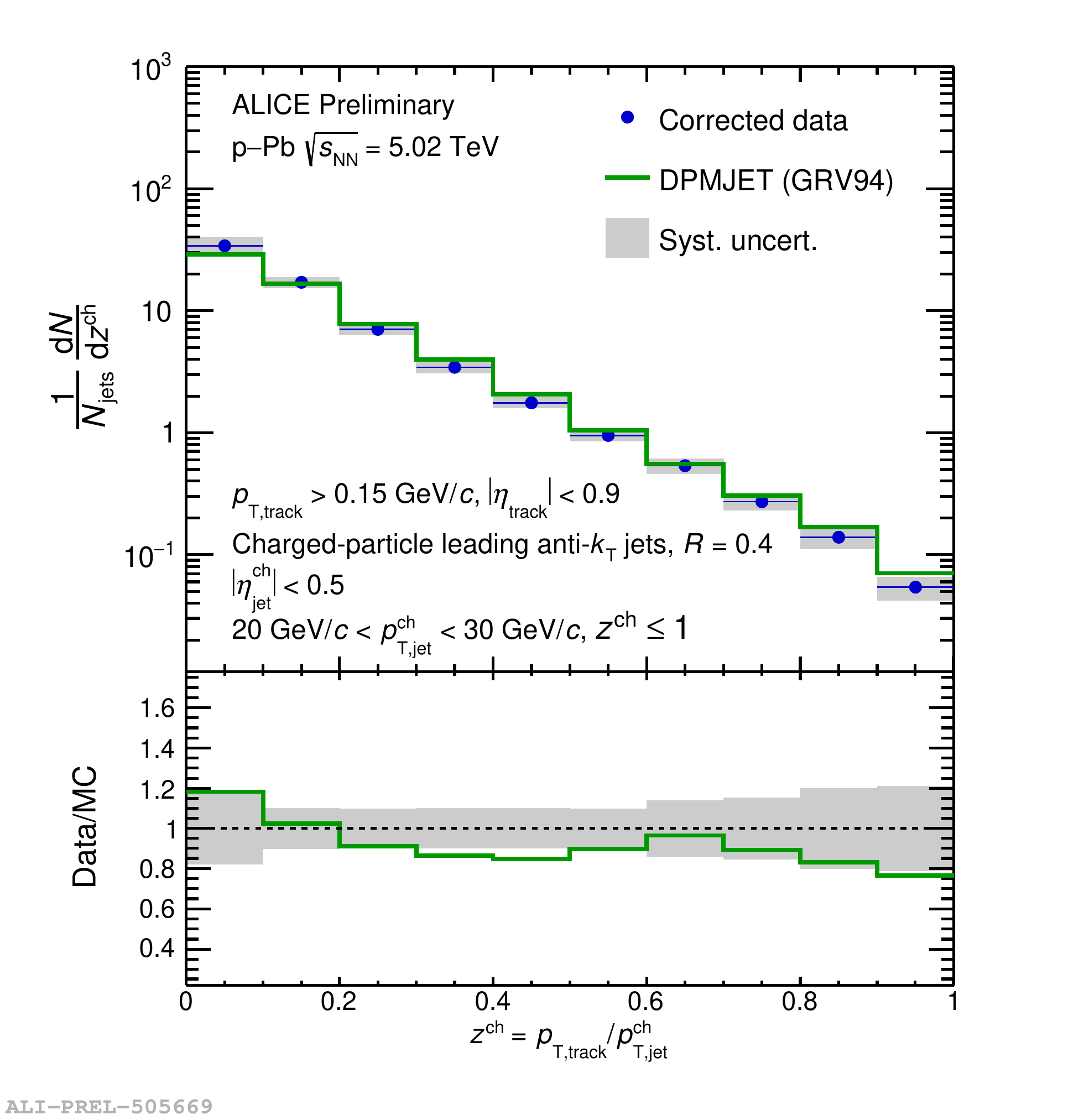}
		\caption{Top panel: Unfolded and UE subtracted $z^{\rm ch}$ distribution for leading jets with 20 $< p_{\rm T, jet}^{\rm ch} <$ 30 GeV/\textit{c}. Bottom panel: ratio between the data and MC results.}
		\label{Fig:FFMB20to30}
	\end{subfigure}
	\vspace*{-0.6cm}
	\begin{subfigure}[b]{0.4\linewidth}
		\includegraphics[width=0.9\columnwidth]{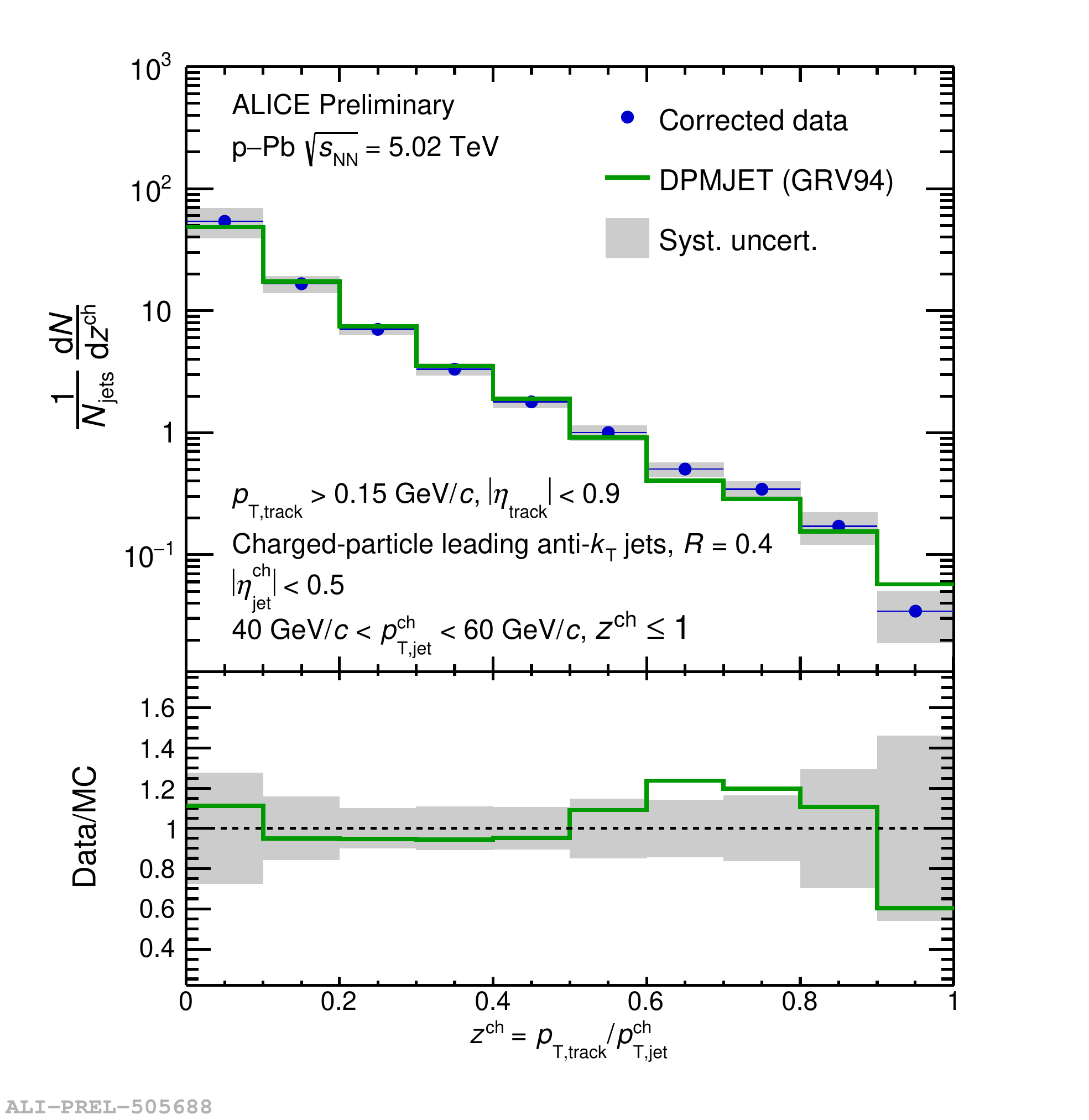}
		\caption{Top panel: Unfolded and UE subtracted $z^{\rm ch}$ distribution for leading jets with 40 $< p_{\rm T, jet}^{\rm ch} <$ 60 GeV/\textit{c}. Bottom panel: ratio between the data and MC results.}
		\label{Fig:FFMB40to60}
	\end{subfigure}
	\hspace*{1.9cm}	
	\begin{subfigure}[b]{0.43\linewidth}
		\includegraphics[width=0.87\columnwidth]{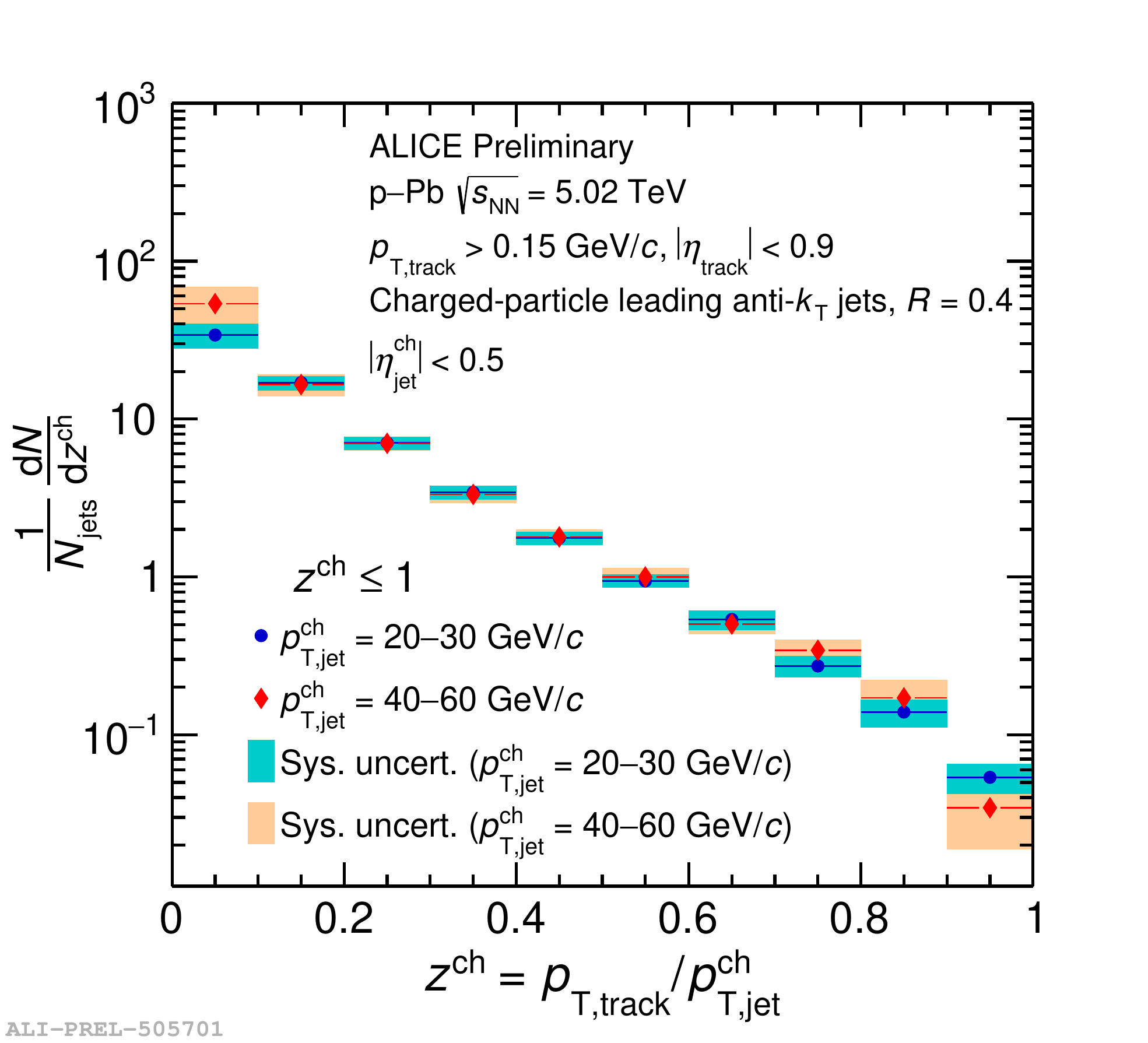}
		\caption{Unfolded and UE subtracted $z^{\rm ch}$ distributions for leading jets with 20 $< p_{\rm T, jet}^{\rm ch} <$ 30 GeV/\textit{c} and 40 $< p_{\rm T, jet}^{\rm ch} <$ 60 GeV/\textit{c}. Scaling of jet fragmentation with $p_{\rm T, jet}^{\rm ch}$ is observed within systematic uncertainties.}
		\label{Fig:FFAllMB}
	\end{subfigure}
	\caption{}
\end{figure}

Figure~\ref{Fig:NchMB} (top panel) shows the $\langle N_{\rm ch}\rangle$ distribution as a function of leading jet $p_{\rm T, jet}^{\rm ch}$. The blue markers represent the data and the green curve shows the DPMJET (GRV94~\cite{GRV94}) prediction. The ratio between the data and DPMJET is depicted in the bottom panel of Figure~\ref{Fig:NchMB}. It is observed that $\langle N_{\rm ch}\rangle$ increases with $p_{\rm T, jet}^{\rm ch}$ and DPMJET reproduces the data for $p_{\rm T, jet}^{\rm ch}>$ 30 GeV/$c$
within uncertainties.
Figures~\ref{Fig:FFMB20to30} and ~\ref{Fig:FFMB40to60} (top panels) show the $z^{\rm ch}$ distributions for leading jets with 20 $< p_{\rm T, jet}^{\rm ch} <$ 30 GeV/\textit{c} and 40 $< p_{\rm T, jet}^{\rm ch} <$ 60 GeV/\textit{c} respectively. Results are compared with the DPMJET (GRV94) predictions. The ratios between the data and DPMJET predictions are shown in the bottom panels. It is found that DPMJET reproduces the $z^{\rm ch}$ distributions in both $p_{\rm T, jet}^{\rm ch}$ ranges within systematic uncertainties.
The comparison of the $z^{\rm ch}$ distributions in the intervals 20 $< p_{\rm T, jet}^{\rm ch} <$ 30 GeV/\textit{c} and 40 $< p_{\rm T, jet}^{\rm ch} <$ 60 GeV/\textit{c} shown in Figure~\ref{Fig:FFAllMB} indicate that the $z^{\rm ch}$ distribution follows a scaling behaviour with $p_{\rm T, jet}^{\rm ch}$ within systematic uncertainties.
\section{Summary}
We have presented the measurement of charged-particle jet properties in minimum bias p--Pb collisions at 5.02 TeV in ALICE. Results are compared with DPMJET (GRV94) predictions, which reproduces both the measured distributions ($\langle N_{\rm ch}\rangle$ and $z^{\rm ch}$) within uncertainties except for $\langle N_{\rm ch}\rangle$ at very low $p_{\rm T, jet}^{\rm ch}$ ($< 30$ GeV/$c$). A scaling of jet fragmentation with leading charged-particle jet $p_{\rm T}$ is observed within systematic uncertainties.

\section*{Acknowledgements}
Prottoy Das would like to acknowledge the Institutional Fellowship of Bose Institute, the ALICE project grant [SR/MF/PS-02/2021-BI (E-37125)] and the computing server facility at Bose Institute.

\end{document}